# Cosmological parameters and the redshift distribution of gravitational lenses

**Phillip Helbig**⋆ **and Rainer Kayser**⋆⋆

Hamburger Sternwarte, Gojenbergsweg 112, D-21029 Hamburg-Bergedorf, Germany

**Abstract.** For known gravitational lens systems the redshift distribution of the lenses is compared with theoretical expectations for $10^4$ Friedmann-Lemaître cosmological models, which more than cover the range of possible cases. The comparison is used for assigning a relative probability to each of the models. The entire procedure is repeated for different values of the inhomogeneity parameter $\eta$ and the limiting spectroscopic magnitude $m_{\rm lim}$, which is important for selection effects. The dependence on these two parameters is examined in more detail for the special cases $\lambda_0 = 0$ and $k = 0$.

Previous results that this method is a better probe for $\lambda_0$ than $\Omega_0$ are confirmed, but it appears that the low probability of models with large $\lambda_0$ values found using similar methods is due to a selection effect.

The power of this method to discriminate between cosmological models can of course be improved if more gravitational lens systems are found. However, our numerical simulations indicate that a reasonable number of observed systems cannot deliver interesting constraints on the cosmological parameters.

## 1. Introduction

It has recently been suggested by many authors (see, for example, Fukugita et al. (1992) and references therein) that gravitational lensing statistics can provide a means of distinguishing between different cosmological models, most effectively concerning the value of the cosmological constant. This is fortunate, since most of the classical methods for determining cosmological parameters are more sensitive to other quantities such as the density ($\Omega_0$) or deceleration ($q_0$) parameters. It has even been suggested (Carroll et al. 1992) that gravitational lens statistics based on *current* observations *already* give the best upper limits on $\lambda_0$ for world models with $k > 0$, and are the most promising method of doing so for $k = 0$.

Kochanek (1992) has suggested a method based not on the total number of lens systems but rather on the redshift distribution of known lens systems characterised by observables such as redshift and image separation. Looking at a few different models, he concludes that flat, $\lambda$-dominated models are five to ten times less probable than more 'standard' models. The advantage of this method is that it is not plagued by normalisation difficulties as are most schemes involving the total number of lenses. The aim of this paper is to extend this method to arbitrary Friedmann-Lemaître cosmological models as well as to look at the influence of observational bias concerning the brightness of the lens. In addition, numerical simulations are used to estimate the usefulness of the method when more systems are available.

It is important to note that the method described in this paper treats $\lambda_0$ and $\Omega_0$ as independent parameters, that is, they can in principle be determined simultaneously. Also important is the fact that $\Omega_0$ is the *global* value, i.e., determined by the contribution of all components, regardless of degree of homogeneity and so on. This is because the cosmological parameters make themselves felt through the cosmological model; most methods of determining $\Omega_0$ will miss any matter homogeneously distributed on a scale larger than that surveyed.

## 2. Theory

We make the 'standard assumptions' that the Universe can be described by the Robertson-Walker metric and that lens galaxies can be modelled as non-evolving singular isothermal spheres (SIS). If one drops the first assumption, the cosmological parameters $\lambda_0$, $\Omega_0$ and $H_0$ lose their significance; the second assumption allows easy calculation but, more importantly, is probably justified within

*Send offprint requests to*: P. Helbig
⋆ phelbig@hs.uni-hamburg.de
⋆⋆ rkayser@hs.uni-hamburg.de

the attainable accuracy (see Krauss & White (1992) for a discussion). In order to have a well-defined statistical quantity, which is based on the optical depth $d\tau$ for 'strong' lensing events,[1] this discussion is limited to gravitational lens systems with sources which are multiply imaged ($\rightarrow$ image separation) by isolated ($\rightarrow$ negligible cluster influence) single galaxies and with known source and lens redshifts. An additional requirement is that the system must have been found without any biases concerning the redshift of the lens. (See Kochanek (1992) for a discussion of these selection criteria.)

Making use of the fact that the SIS produces a constant deflection angle, i.e., independent of the position of the source with respect to the optical axis (defined as passing through observer and lens), one can define the angular cross section $\pi a^2$ of a *single* lens for 'strong' lensing events (Turner et al. 1984):

$$\pi a^2 = 16\pi^3 \left(\frac{v}{c}\right)^4 \left(\frac{D_{\rm ds}}{D_{\rm s}}\right)^2 , \qquad (1)$$

where $v$ is the one-dimensional velocity dispersion of the lens galaxy, $c$ the speed of light and $D_{\rm ds}$ ($D_{\rm s}$) the angular size distance between lens and source (observer and source). Following Kochanek (1992), one can arrive at an expression for the optical depth as follows.

For a *fixed mass* and mass distribution ($\rightarrow v$), world model and $z_{\rm s}$, the *differential* optical depth due to all lenses of a given mass as a function of $z_{\rm d}$ is of course proportional to the number of lenses of the given mass per $z_{\rm d}$-interval and to the cross section for strong lensing events. In order to arrive at an expression for $d\tau$ for a *fixed image separation*, one needs to know the relative number of lenses which, under the given circumstances, can produce this image separation. This can be done by using the Schechter luminosity function (Schechter 1976) as well as the Faber-Jackson and Tully-Fisher relations (Faber & Jackson 1976, Tully & Fisher 1977), which give the dependence of the velocity dispersion on the luminosity for elliptical and spiral galaxies, respectively. Bringing in the familiar parameters and dropping all terms which are concerned only with normalisation, one arrives, after some tedious but trivial calculations, at the expression

$$\frac{d\tau}{dz_{\rm d}} = (1+z_{\rm d})^2 \frac{a}{a*}\frac{\gamma}{2}\left(\frac{a}{a*}\frac{D_{\rm s}}{D_{\rm ds}}\right)^{\frac{\gamma}{2}(1+\alpha)} \times$$
$$D_{\rm d}^2 \frac{1}{\sqrt{Q(z_{\rm d})}} \exp\left(-\left(\frac{a}{a*}\frac{D_{\rm s}}{D_{\rm ds}}\right)^{\frac{\gamma}{2}}\right) , \qquad (2)$$

where $a* := 4\pi\left(\frac{v*}{c}\right)$ ($v* := v$ of an $L*$ galaxy), $\gamma$ is the Faber-Jackson/Tully-Fisher exponent, $\alpha$ the Schechter exponent, $D_{\rm d}$ the angular size distance between the observer and the lens and

$$Q(z_{\rm d}) := (1+z_{\rm d})^2(\Omega_0 z_{\rm d} + 1 - \lambda_0) + \lambda_0 . \qquad (3)$$

[1] See Schneider et al. (1992) for a clarification of the concept of optical depth in lensing.

Equation (2) is independent of the Hubble constant since the dependences on $H_0$ in the angular size distances and in the Faber-Jackson/Tully-Fisher relation cancel. In order to facilitate comparison with other authors, we have chosen the 'standard values' $-1.1$, 2.6, 4, 144 km/s and 276 km/s for the Schechter exponent, the Tully-Fisher exponent, the Faber-Jackson exponent, $v*_{\rm spiral}$ and $v*_{\rm elliptical}$, respectively. (The value for $v*_{\rm elliptical}$ includes the factor $(3/2)^{\frac{1}{2}}$ advocated by Turner et al. (1984) and so our elliptical galaxies correspond to the $c = 2$ models examined by Kochanek (1992).)

The optical depth depends on the cosmological model through $Q(z_{\rm d})$ as well as through the angular size distances, because of the fact that $D_{ij} = D_{ij}(z_i, z_j; \lambda_0, \Omega_0, \eta)$. The influence of $\eta$, which gives the fraction of homogeneously distributed, as opposed to compact, matter is felt only in the calculation of the angular size distances, whereas the cosmological model in the narrower sense makes its influence felt here as well as through $Q(z_{\rm d})$.

In general, there is no analytic expression for the $D_{ij}$; they can be obtained by the solution of a second-order differential equation. (See Kayser (1985) for the derivation of the differential equation, also Linder (1988) for a more general formulation. For an equivalent derivation for $\lambda_0 = 0$ see Schneider et al. (1992). Kayser et al. (1995) give a general discussion and an easy-to-use numerical implementation.) If one has an efficient method of calculating the angular size distances, it is easy to evaluate Eq. (2) for various world models described by the parameters $\lambda_0$, $\Omega_0$ and $\eta$.

Worthy of note is the *independence* of Eq. (2) on the source luminosity function (which of course will generally itself depend on $z_{\rm s}$ as well), the relative numbers of galaxy types (the galaxy type for a particular lens is known) and the fraction of galaxies in clusters (the method looks only at field galaxies); these factors have to be taken into account when doing statistics based on the total number of lenses. Also, Eq. (2) is insensitive to finer points of the mass model such as core radius and ellipticity (cf. Krauss & White (1992), Narayan & Wallington (1992)). The main idea is to compare the observed distribution of lens redshifts with theoretical expectations for various world models; the method is described in the next section.

Equation 2 can take on appreciable values at intermediate redshifts *even though the lens galaxy would be too faint to be seen at the redshift in question* ($m > m_{\rm lim}$). In order to correct for this effect, we have calculated the redshift at which the lens galaxy would become too faint to have its redshift measured for the investigated cosmological model and truncated $d\tau$ at this point. (Details in the next section.) It is immediately obvious that failure to correct for the faintness of the lens galaxies will artificially exclude cosmological models with a high median redshift in Eq. (2), which might otherwise not be excluded.

**Table 1.** Gravitational lens systems used. For references see Refsdal & Surdej (1994) and references therein. Note that $\theta''$ corresponds to the radius of the Einstein ring or *half* the image separation

| name | images | $\theta''$ | source | $m_{\rm source}$ | $z_{\rm s}$ | lens | $m_{\rm lens}$ | $z_{\rm l}$ | comments |
|---|---|---|---|---|---|---|---|---|---|
| 0142-100 | 2 | 1.1 | QSO | B = 17.0<br>B = 19.1 | 2.719 | EG | R = 19.0 | 0.493 | 'typical' multiply imaged QSO |
| 0218+357 | 2 + ring | 0.165 | radio lobe | | 0.94 | SG | r ≈ 20 | 0.6847 | 'radio ring' |
| 1115+080 | 4 | 1.15 | QSO | B = 17.2<br>B = 17.2<br>B = 18.7<br>B = 18.2 | 1.722 | EG | R = 19.8 | 0.29 | |
| 1131+0456 | 2 + ring | 1.05 | EG, radio lobe | | 1.13 | EG | R = 22 | 0.85 | 'radio ring' |
| 1654+1346 | ring | 1.05 | radio lobe | | 1.74 | EG | R = 18.7 | 0.254 | 'radio ring' |
| 3C324 | 3 | 1.0 | AGN | R = 22.7<br>R = 23.3 | 1.206 | SG | R = 22.5 | 0.845 | |

## 3. Calculations

The following gravitational lens systems meet our selection criteria: 0142-100 (= UM 673 ), 0218+357, 1115+080 (= Triple Quasar), 1131+0456, 1654+1346 and 3C324. (See Table 1 for observational data on these systems.) We considered the following ranges of values for the cosmological parameters:

$$\begin{aligned} -10 &< \lambda_0 < +10 \\ 0 &< \Omega_0 < 10 \\ 0 &< \eta < 1 \end{aligned}$$

which, of course, are much larger than contemporary knowledge demands. However, the history of cosmology shows that the knowledge of today is often out of fashion tomorrow, so that we prefer to develop an approach capable of dealing with a wide range of cosmological models. Also important is the fact that it would be an additional, though by no means necessary, point in favour of the validity of the method if it assigns the highest probability to a cosmological model within the presently accepted canonical parameter space.

We looked at 100 × 100 models in the $\lambda_0$-$\Omega_0$ plane for

$$\eta = 0.0,\ 0.3,\ 0.5,\ 0.7,\ 1.0 \text{ for } m_{\rm lim} = 23.5$$

and

$$m_{\rm lim} = 23.5,\ 24.5,\ \infty \text{ for } \eta = 0.5$$

(Johnson $R$ magnitudes). In addition, we looked at 100 × 100 models in the $\eta$-$\Omega_0$ and $m_{\rm lim}$-$\Omega_0$ planes for the special cases of $\lambda_0 = 0$ and $k = 0$.

Before one can examine the relative probability of a given cosmological model, one must first see if it is compatible with the observations. (Of course, this does not imply that the model is compatible with *all* observations, merely with the ones necessary for this analysis: $z_{\rm l}$, $z_{\rm s}$, $\theta''$ and galaxy type.) One obvious restriction is that the largest source redshift $z_{\rm s,max}$ in the sample must be smaller than $z_{\rm max}$, the maximum redshift possible in the cosmological model in question. (See, e.g., Stabell & Refsdal (1966) or Feige (1992) for a discussion of these cosmological models.) Another restriction concerns the brightness of the lens. From the observables $z_{\rm l}$, $z_{\rm s}$, $\theta''$ and galaxy type one can use Eq. (1) to calculate the velocity dispersion $v$, transform this to an absolute luminosity using the Faber-Jackson or Tully-Fisher relation and then calculate the apparent magnitude for the given cosmological model (given by the angular size distance up to powers of $(1+z_{\rm l})$ and $K$-corrections[2]). If this calculated lens brightness for at least one lens is fainter than $m_{\rm lim}$ then the model is also incompatible with the observations. Since a realistic value for $m_{\rm lim}$ is at least a magnitude fainter than any $m_{\rm lens}$ value in Table 1, there is no danger that the actual cosmological model would be excluded by this restriction, even allowing for the uncertainty in calculating the lens brightness in the matter described.

[2] The apparent luminosity of the lens galaxy was calculated for the Johnson $R$-band using the $K$-corrections of Coleman, Wu & Weedman (1980) and applying a standard $B-R$ correction (since the $B$-band Faber-Jackson and Tully-Fisher relations were used). Since these $K$-corrections are based on displacement of standard spectra at $z = 0$ which extend into the UV-band, they are given only up to $z = 2.0$, where evolutionary effects would in any case have to be considered. However,

In these two cases we assigned the corresponding world model a probability of 0 in our plots, indicated by white. Since the value 0 doesn't occur otherwise, all white areas are due to these two restrictions. Otherwise, to measure the relative probability of a given cosmological model, we defined the quantity $f$ as follows:

$$0 < f := \frac{\int_0^{z_{\rm l}} {\rm d}\tau}{\int_0^{z_{\rm s}} {\rm d}\tau} < 1, \qquad (4)$$

where $z_{\rm l}$ is the *observed* lens redshift for a particular system. ($z_{\rm d}$ is used to denote the variable corresponding to lens redshift as opposed to the measured value for a particular lens.) The distribution of the different $f$ values (one for each lens system in the sample) in $b$ equally-sized bins in the interval ]0,1[ gives the relative probability $p$ of a given cosmological model, with

$$p = \prod_{i=1}^{b} \frac{1}{n_i!} \qquad (5)$$

where $n_i$ is the number of systems in the $i$-th bin. This definition allows only a few discrete values, of course. The variable $b$ is a free parameter; since the most information is obtained when $b$ is equal to the number of systems, we adopt this value for $b$.

Were the other observables the same for all systems, the redshift distribution should be given by Eq. (2); since this is not the case, the quantity $f$ is defined, which allows one to compare the observed with the expected redshifts for different observables and hence different relative probability curves (Eq. (2)) for each system. Our *ansatz* is thus to expect that the $f$-values should be uniformly distributed for the correct cosmological model (barring intrinsic scatter, of course). Simple combinatorics (the number of ways to distribute $a$ objects in $b$ bins) and neglecting normalisation then leads to Eq. (5). For a given world model, the 6 $f$-values (one for each gravitational lens system used) are calculated, and these values are used to determine the relative probability via Eq. (5).

## 4. Results and discussion

Figure 1 shows the relative probability of several cosmological models as given by Eq. (5). The grey scale in all plots is the same, regardless of the maximum and minimum values of each individual plot. The resolution is 0.2 in $\lambda_0$ and 0.1 in $\Omega_0$, thus giving 10.000 different models. The scale is at the right. (Because the relative probability can only take on a few discrete values, contour levels aren't very useful as indicators of the relative probability, since they would merely indicate the boundaries between areas of constant probability. In order to indicate the values directly, Fig. 1 plots the relative probability on a grey scale. Although in themselves not important, the interested reader can read off the probabilities directly in the legend, where the discrete values which actually appear in the plots have been marked. In addition, Table 2 gives the values of the relative probability which occur in each plot.) All plots except (9), (10), (12) and (13) are in the $\lambda_0$-$\Omega_0$ plane. Plot (8) gives some orientation in this plane. The diagonal line from upper left to lower centre corresponds to $k = 0$; the six curves are, left to right, for $ht_0 = 4$, 5, 6, 8 and 10 $\times 10^9$a ($h := H_0 \cdot 100^{-1}{\rm km}^{-1} \cdot {\rm s} \cdot {\rm Mpc}$) as well as the border to the so called bounce models, i.e. models with no big bang and thus a maximum redshift $z_{\rm max}$; the line on the right corresponds to $q_0 = -5$. The values of the fixed parameters $\eta$ and $m_{\rm lim}$ are indicated on each plot. Although some of the area in this plane is definitely excluded[3] by simple arguments, a probability has been computed for each world model, in keeping with the second of our motivations mentioned above. In the white area at the right the probability is 0 because these world models have a maximum redshift lower than the highest redshift in Table 1; in plots (1), (2), (3) and (7) there is an additional white area ($p = 0$) separated from the first one by a strip where $p > 0$ due to the fact that at least one lens is fainter at its observed redshift than $m_{\rm lim}$. (The brightness of the lens was not used as an additional constraint due mainly to the fact that the computed brightnesses are only correct to about a magnitude or so (Kochanek 1992).)

**Table 2.** Values of the relative probability in the plots in Fig. 1

| $p$ | 1 | 2 | 3 | 4 | 5 | 6 | 7 | 9 | 10 | 11 | 12 | 13 |
|---|---|---|---|---|---|---|---|---|---|---|---|---|
| $\frac{1}{1}$ | | | | | | x | | | | | | |
| $\frac{1}{2}$ | | | | x | x | x | x | | | | | |
| $\frac{1}{4}$ | x | x | x | x | x | | x | x | x | x | | |
| $\frac{1}{6}$ | x | x | x | x | x | x | x | x | x | x | | x |
| $\frac{1}{8}$ | x | x | | | | | | | | | | |
| $\frac{1}{12}$ | x | x | x | x | x | | x | x | x | x | x | x |
| $\frac{1}{24}$ | x | x | x | x | x | | x | | | x | x | |
| $\frac{1}{36}$ | | x | | | x | | x | x | | x | | |
| $\frac{1}{48}$ | x | x | x | x | x | | x | x | | x | x | |

in most cases the galaxy becomes too faint at modest redshifts, so the assumption of no evolution made throughout this paper is probably justified.

[3] Even though not every point in the plane, i.e. every combination of $\lambda_0$ and $\Omega_0$, corresponds to a world model which cannot be excluded by simple arguments, nevertheless the ranges of the individual parameters are allowed assuming the lowest realistic values for $H_0$ and the age of the universe.

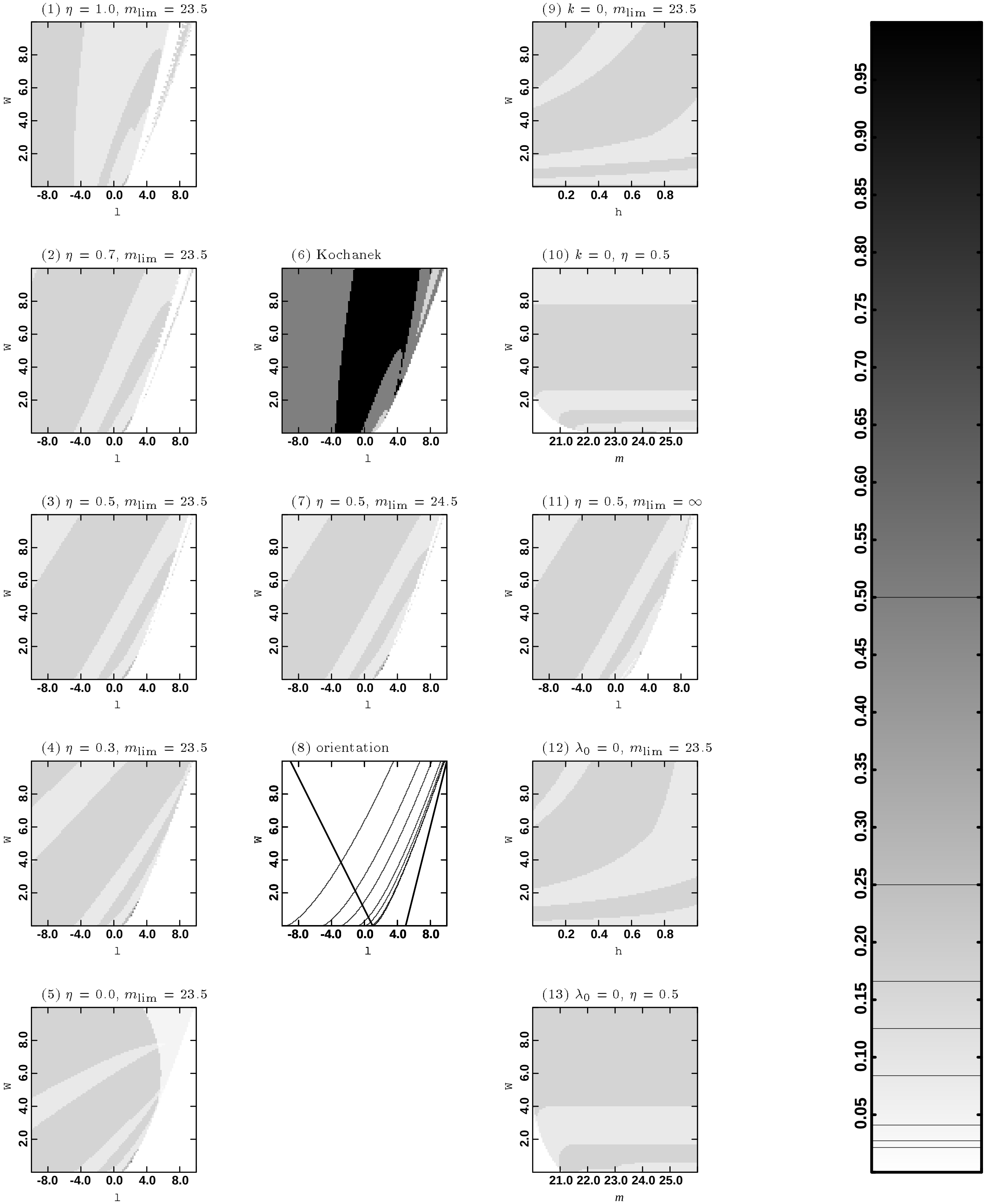

**Fig. 1.** Results for the systems in Table 1. The relative probability for different cosmological models is plotted linearly on the scale shown at the right, where the discrete values which appear in the plots are marked. Fixed parameters are indicated above each plot. For comparison, the results for the systems used in Kochanek (1992) are included in plot (6). Plot (8) is for orientation

Looking at plots (1), (2), (3), (4), (5), (7) and (11) the general impression is that, apart from some small-scale structure at the lower right of the $p > 0$ area which appears to be significant, there is not much structure. Since there are only discrete values of $p$, there are some discontinuities. Apart from this large-scale-structure, however, there is not much change from world model to world model. The curves along which the probability is constant are more vertical than horizontal, suggesting that this method better probes $\lambda_0$ than $\Omega_0$. Although not of any use statistically, for each plot in Fig. 1 the Kolmogorov-Smirnov probability (see Sect. 5) was also computed. In the interest of brevity, we have not included these plots here; however, the structure is qualitatively similar (without the sharp discontinuities, of course) and also hints at the significance of the small-scale structure mentioned above.

Plots (1), (2), (3), (4) and (5) show the effect of varying $\eta$. As one can see, the general structure doesn't change much, especially if one concentrates on the area near the roughly vertical thin strip (which would still be allowed assuming higher values for $H_0$ and/or the age of the universe) for relatively low $\Omega_0$ values ($\Omega_0 < 4$, say). This is compatible with the result of Fukugita et al. (1992) which indicates only a weak dependence of the lensing cross section on $\eta$. $\eta$ enters into the calculation only in the computation of the angular size distances, whereas $\lambda_0$ and $\Omega_0$ also enter the calculation through the function $Q(z_{\rm d})$ as defined in Eq. (3). Also, the particular combination of the angular size distances involved in computing the lensing cross section, $(D_{\rm d}D_{\rm ds})/D_{\rm s}$ (see, e.g., Fukugita et al. (1992) or Kochanek (1992); in Eq. (2) the explicit dependence on $D_{\rm d}$ has been lost in the necessary variable transformation and integration (cf. Kochanek (1992))), is relatively insensitive to $\eta$. This is not true, for example, for the combination $D_{\rm ds}/(D_{\rm d}D_{\rm s})$, which is important for computing $H_0$ from the time delay between the different images of a multiply imaged source (cf. Fukugita et al. (1992)).

Plots (3), (7) and (11) show the effect of varying $m_{\rm lim}$. Of course, the additional white area described above becomes smaller as $m_{\rm lim}$ becomes fainter, disappearing for $m_{\rm lim} \rightarrow \infty$ (this condition is sufficient but not necessary for every lens galaxy to be brighter than $m_{\rm lim}$ for the world models examined). The main differences, however, occur in the small-scale structure at the lower right of the $p > 0$ area: the fainter $m_{\rm lim}$ becomes, the less probable the models near the border of this area appear. For example, if one compares the models near the Einstein-de Sitter model ($\lambda_0 = 0$ and $\Omega_0 = 1$) with the models near the de Sitter model ($\lambda_0 = 1$ and $\Omega_0 = 0$), two models which have been examined rather extensively in the literature on lensing statistics, especially in the direction 'perpendicular' to the curves dividing different probabilities, then one sees that for $m_{\rm lim} = 23.5$ and $m_{\rm lim} = 24.5$ those near the de Sitter models have a roughly equal but slightly higher relative probability than those near the Einstein-de Sitter model; only for $m_{\rm lim} = \infty$ is the situation reversed, those near the Einstein-de Sitter model having a clearly higher probability.

This is easy to understand, since it is these models near the de Sitter model which have a maximum in $d\tau$ at relatively large redshifts (cf. Kochanek (1992)); for realistic values of $m_{\rm lim}$, one cannot see the lens galaxies at these redshifts. If one uses a realistic value for $m_{\rm lim}$, one compares the redshift distributions for the different world models (given by Eq. (2)) at small redshifts, where they don't differ very much. If one takes $m_{\rm lim} = \infty$, implying that one could measure the redshifts of the lenses at all redshifts, whatever their brightness, then it appears that models with a large median redshift in Eq. (2), such as those on the right hand border of the $p > 0$ area including the models with a large cosmological constant examined by Kochanek (1992), are improbable. However, this is merely a selection effect. *The relatively low lens redshifts in Table 1 don't mean that world models with a large median lens redshift are improbable; it means that we can't see the lenses at these redshifts. Comparing probability distributions assuming that we could artificially excludes these models in preference to models like the Einstein-de Sitter model with a small median lens redshift.* Introducing $m_{\rm lim}$ makes the situation more realistic, but means comparing the distributions at small redshifts. Thus, the power to discriminate between various world models is reduced.

For realistic values of $m_{\rm lim}$, such as those in plots (3) and (7), the exact value of $m_{\rm lim}$ isn't very important; what does make a difference is assuming a value which is much too faint. This is easy to understand, since, near the redshift where a lens galaxy typically becomes fainter than a realistic value for $m_{\rm lim}$, the function $m(z_{\rm d})$ is rather steep. This means that a change in $m_{\rm lim}$ by a magnitude or two corresponds to a relatively small change in $z_{\rm d}$ and thus to a correspondingly small change in the area under the $d\tau$ curve up to this $z_{\rm d}$ value; thus, there is little influence on the value of $f$ as defined in Eq. (4). This is illustrated in Fig. 2.

For comparison, we have also tested the method on the systems used by Kochanek (1992), using $m_{\rm lim} = \infty$ und $\eta = 1$, both of which he implicitly assumes.[4] The results are in plot (6) where the relative probabilities are $\frac{1}{6}$, $\frac{1}{2}$ and 1 and comparing the various models examined by Kochanek confirm his conclusions. For instance, the relative probabilities of the Einstein-de Sitter and de Sitter model are 1 and $\frac{1}{6}$, confirming his result that flat, $\lambda$-dominated models are 5–10 times less probable than standard ones. (However, taking $m_{\rm lim}$ into account and/or us-ing only directly measured lens redshifts would produce quite different results, as discussed above.) This plot artificially indicates a low probability for models near the de Sitter model for the same reasons as those discussed in connection with plot (11).

[4] Of course, when one considers finite values for $m_{\rm lim}$, one cannot include systems with lens redshifts which have been determined by means other than measured emission redshifts, such as absorption lines (which assumes that the lens is also the absorber).

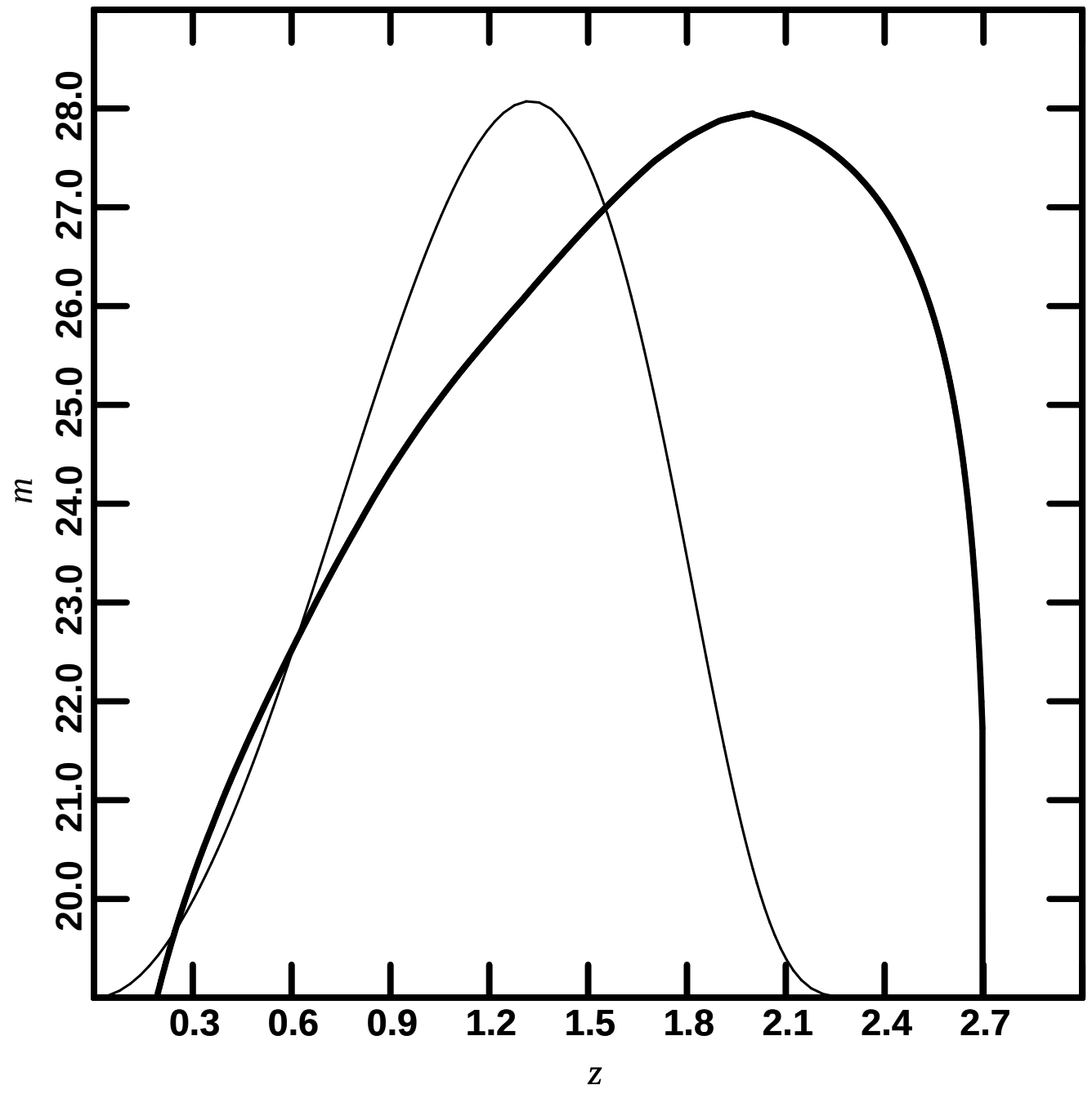

**Fig. 2.** The relative differential optical depth (thin curve) and the calculated lens brightness $m$ (thick curve) as functions of $z_d$. The world model is the de Sitter model ($\lambda_0 = 1.0$ $\Omega_0 = 0.0$; the value of $\eta$ doesn't matter since there is no matter) and the observables are those for the gravitational lens system $0142-100$ (see Table 1). The ordinate gives the magnitude in Johnson $R$

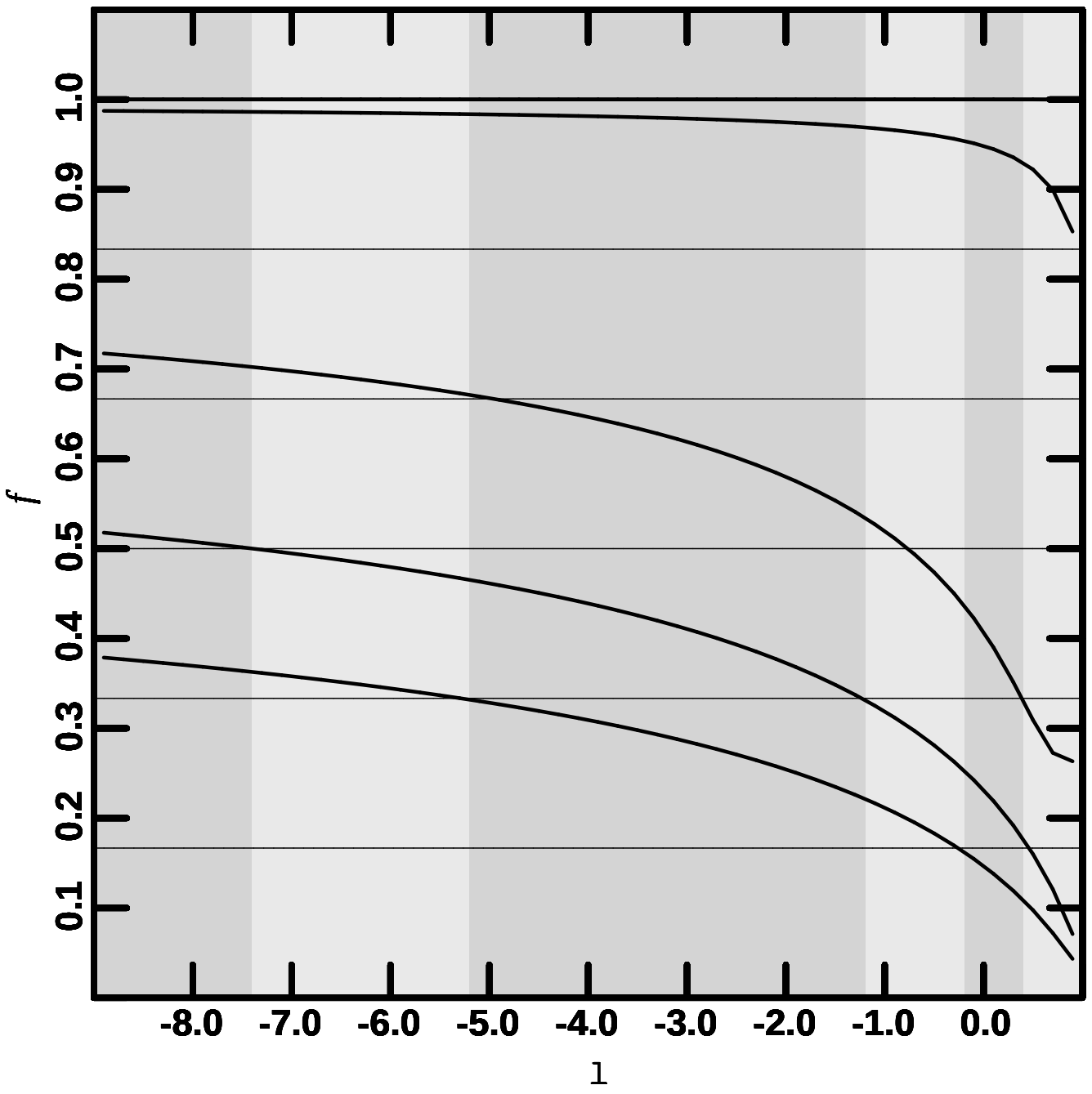

**Fig. 3.** 'Oscillations' in the relative probability. The thin horizontal lines are the bin boundaries. The thick curves show, from top to bottom, the $f$-values given by Eq. (4) for the gravitational lens systems 1131+0456, 3C 324, 0218+357, $0142-100$, $1115+080$ and $1654+1346$, respectively. Since two systems have such large lens redshifts that $d\tau$ is practically $= 0$ at the corresponding redshift, their $f$-values (in these world models) are practically $= 1$ and cannot be distinguished from each other or from the bin boundary at $f = 1$ in the resolution of the plot. Table 1 shows that these are the two highest lens redshifts in the sample. Beneath the lines and curves, shown in the same way as in Fig. 1, is the corresponding relative probability

Plots (9) and (10) examine the influence of $\eta$ and $m_{\rm lim}$, respectively, for the special case $k = 0$. Plots (12) and (13) examine the influence of $\eta$ and $m_{\rm lim}$, respectively, for the special case $\lambda_0 = 0$. In plots (9) and (12) one can easily see the weak dependence on $\eta$, especially for small values of $\Omega_0$. In plots (10) and (13) one can see the even weaker dependence on $m_{\rm lim}$ in this range. There are hints toward fainter magnitudes of a declining probability for small $\Omega_0$ values, as discussed above. The white area at the left in plots (10) and (13) is due to the fact that at least one lens is fainter at its observed redshift than the corresponding $m_{\rm lim}$ value. (Since these $m_{\rm lim}$ values are unrealistically small, these models are not incompatible with the observations.)

At first glance, the 'oscillations' in the relative probability might appear somewhat puzzling. According to Eq. (5), a higher probability is obtained for a more regular distribution. Since Eq. (5) only allows *discrete values* for the relative probability, a 'jump' occurs when the number of systems in a certain bin changes (unless offset by a corresponding change in another bin). This can be seen in Fig. 3, where for demonstration purposes the $f$-values given by Eq. (4) for each gravitational lens system used are plotted as functions of $\lambda_0$ for $k = 0$ and $\eta = 0.3$. (That is, for the world models along the $k = 0$ line in plot (4) in Fig (1).) It can be seen that, although—due to the definition—the relative probability changes by noticeable amounts between a few discrete values, nevertheless the $f$-values themselves are smoothly varying functions of the world model. Figure 3 also makes the following general conclusions clear in this specific example.

- The fact that a couple of systems have $f$-values which are practically $= 1$ limits the maximum probability, since these are always in the same bin.
- 'Oscillations' between different probabilities have no physical significance, but rather are merely artifacts of the particular lens redshifts. On the other hand, extremely low probabilities, e.g. all six systems in the same bin ($p = \frac{1}{720}$), would be more indicative of a low-probability cosmological model.
- Apart from the oscillations, a trend (the $f$-values increasing to the left in Fig. 1) can be seen which would

indicate a probability low enough to reject the corresponding cosmological models, were the parameter space examined larger. Thus, the method might be able to exclude 'extreme' cosmological models.

The interested reader can use Fig. 3 together with Eq. (5) to see how the relative probability is arrived at.

The statistical significance of all results in this section is discussed in the next section.

## 5. Numerical simulations

Numerical simulations were done for $m_{\rm lim} = 23.5$ and for $\eta = 1$. This value of $m_{\rm lim} = 23.5$ is the most realistic based on the present state of observations and using only *one* value for $m_{\rm lim} = 23.5$ as well as for $\eta$ is justified based on the weak dependence of the results on these parameters, as discussed in Sect. 4.

The observables $\theta''$ (the radius of the Einstein ring or *half* the image separation corresponding to the diameter of the Einstein ring), $z_{\rm s}$ and galaxy type were chosen randomly from an interval roughly corresponding to the observed range of values in order to produce synthetic data comparable to real observations. It is important to note that neither the exact range nor the shape of the distribution matters, since the method looks at the redshift distribution of the lenses with the other parameters fixed by observation. For convenience, a flat distribution was chosen for each of the observables. For a given cosmological model, the corresponding lens redshift $z_{\rm l}$ for each system was calculated from the observables and a randomly generated $f$ through (numerical) inversion of Eq. (4). This catalog was then used to determine a relative probability for each of the points in the $\lambda_0$-$\Omega_0$ plane in the same manner as for the real systems.

With the probability given by Eq. (5), based on simple combinatorics, one cannot know to what degree the values for each cosmological model are influenced by statistical fluctuations. However, with such a small number of systems, there is really no other method of computing a relative probability. We expect that the distribution of the $f$-values, barring statistical fluctuations, should be flat for the correct cosmological model. So we need a test to compare this distribution with a flat probability distribution. The Kolmogorov-Smirnov (K-S) test is of course a well understood method for testing if two distributions are statistically significantly different. (See, e.g., Press et.al. (1992) for a general discussion and definition of the K-S probability.) However, this test can only be used for distributions with more than $\approx 20$ data points. For purposes of comparison, for the systems in Table 1 not only was the probability defined in Eq. (5) computed (shown in Fig. 1) but also the K-S probability. The K-S probability should of course not be taken seriously if there are too few systems.

We have done simulations for a variety of world models and also for numbers of systems between 20 and 50. In the interest of brevity, we present only one plot. Figure 4 shows the results derived from a catalogue of simulated gravitational lens systems. Since, even with 50 systems, no area can be excluded based on the K-S probability – the white area has $p = 0$ due to the fact that at least one lens would be fainter than $m_{\rm lim}$ in these world models, as discussed above – we conclude that, although one can qualitatively understand the physics which at least in part is responsible for the results presented in Fig. 1, the actual relative probabilities are more indicative of intrinsic scatter in the redshifts of the lenses than a hint of the correct cosmological model.

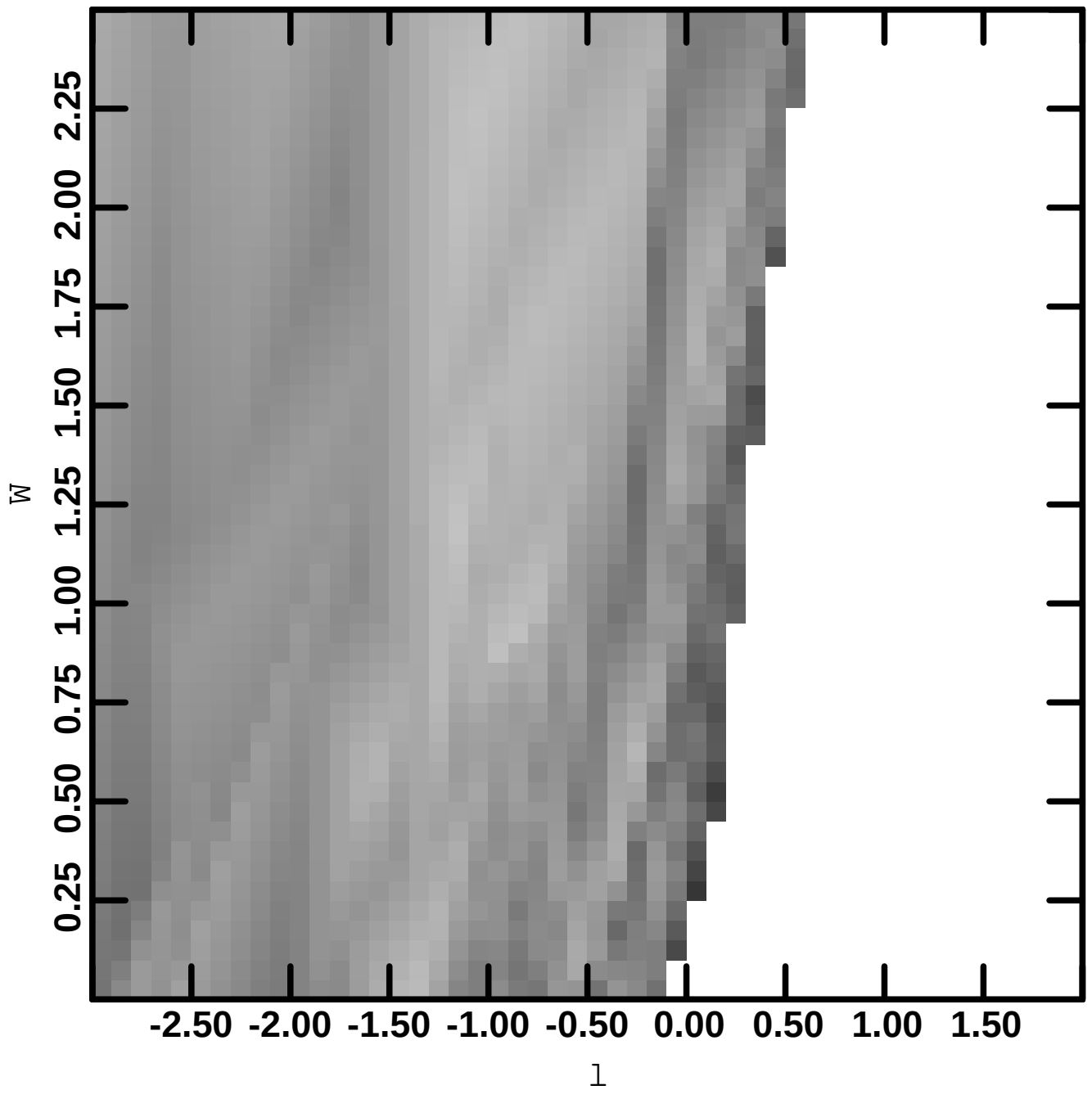

**Fig. 4.** Results based on a catalogue of 50 simulated systems. (Note the different scale on the axes.) The cosmological model used to generate the lens redshifts is the homogeneous Einstein-de Sitter model ($\eta = 1.0$, $\lambda_0 = 0.0$, $\Omega_0 = 1.0$). The grey scale is as in Fig. 1

## 6. Summary and conclusions

In this paper we have extended the method originally proposed by Kochanek (1992) for using the redshift distribution of gravitational lenses to learn something about the cosmological model. This method is particularly attractive since it avoids the normalisation difficulties normally associated with lensing statistics. First, we made use of the equation derived by Kayser (1985; see also Kayser et al. 1995) to be able to examine cosmological models described by arbitrary values of $\lambda_0$, $\Omega_0$ and $\eta$. Second, we looked at the influence of observational bias by introducing the parameter $m_{\rm lim}$. (This means that we cannot include systems where the lens redshift has been estimated by some other means than an emission spectrum.)

Third, we used a more quantitative statistic to look at relative probabilities.

Since the potential reward from using the Kochanek formalism (setting limits on cosmological parameters better than with other methods using only observable quantities and standard assumptions whose validity in this context is undisputed) seemed large, our basic idea was to extend this formalism to enable it to look at a larger number of cosmological models (arbitrary values for $\Omega_0$ and $\lambda_0$ as well as $\eta$) while correcting an obvious limitation (selection effects due to the brightness of the lens) and using a variation of his statistic (Kochanek (1992) basically uses our statistic with two bins) which allows more information about the distribution of the redshift values to be used. That is, we intended to follow the formalism of Kochanek (1992) as closely as possible. Unfortunately, the fact that the relative optical depth given by Eq. (2) is appreciably different only for those cosmological models in which this difference cannot be seen due to the selection effect renders the technique less useful than we had hoped.

We saw that little extra uncertainty in the derived values for $\lambda_0$ and $\Omega_0$ is introduced by letting $\eta$ be a free parameter. The same is true of $m_{\rm lim}$ with the exception that values which are unrealistically faint distort the results. A comparison with the results of Kochanek (1992), confirmed with our formalism, also show the consequences of neglecting $m_{\rm lim}$.

An interesting result is the degeneracy of the derived relative probability along curves roughly parallel to curves of constant world age in the $\lambda_0$-$\Omega_0$ plane. This indicates that the method is more sensitive to $\lambda_0$ than to $\Omega_0$, but also shows that demonstrating the consistency of a given cosmological model with the observations using this statistical method also implies consistency with a large range of other cosmological models.

The dramatic difference caused by not neglecting $m_{\rm lim}$ casts doubt on the degree to which *present* observations, based *only* on the redshift distribution,[5] are able to rule out certain cosmological models; in particular, flat models with a large cosmological constant, having a high median expected lens redshift, become more probable through introducing $m_{\rm lim}$.

Nevertheless, plots (6) and (11) gives an idea of what could be done, if one were able to measure the redshifts of the faintest lens galaxies. For a given image separation, the calculated brightness of the lens galaxy has a minimum at some intermediate redshift; this is typically at about $30^{\rm m}$ in $R$, so that larger telescopes and advances in image processing will probably be able to make some progress on this front in the next several years. If one were able to measure the lens redshift at the minimum brightness, this would have the side-effect of eradicating the dependence on $m_{\rm lim}$. On the other hand, probably more would be lost than gained, because it would no longer be possible to neglect evolutionary effects. For this reason, one could suppose that more progress in the immediate future (barring a revolution in the understanding of evolutionary effects) will probably come from increasing the number of usable systems (through the discovery of more systems and/or through measuring more redshifts in known systems) than from pushing $m_{\rm lim}$ to fainter values.

The results of using the method on synthetic data, however, cast doubt on the statistical significance of our results and on the hope of using this method to exclude certain cosmological models, especially those which cannot be excluded by other tests. Thus, being conservative in our appraisal of what the statistics of redshifts of gravitational lenses can tell us, we conclude that at present and in the foreseeable future this method will probably not give us any useful information.

[5] More information is available in theory by looking at not only the redshift distribution, i.e., the shape of the curve, but also the number of lenses, i.e., the area under the curve. This, however, introduces additional uncertainties due to normalisation. Additionally, one could consider the completeness of the sample, as pointed out to the authors by the referees. For example, the number of lens systems with unmeasured redshifts could in principle be used to exclude cosmological models in which the redshifts could have been measured, or the other way around. This requires information regarding the *reasons* as to why the redshifts haven't been measured—the effects of the cosmology should be separated from the current observational stand, which requires a detailed analysis of the observational literature, or a separate observational programme, both of which are beyond the scope of this paper.

*Acknowledgements.* It is a pleasure to thank S. Refsdal and T. Schramm for helpful discussions.